\documentclass[pral,onecolumn,floatfix,superscriptaddress]{revtex4}
\usepackage{pstricks,graphicx,amsmath,bbm,mathrsfs,amssymb,psfrag,pifont,times,mathptmx}
\usepackage[utf8]{inputenc}
\usepackage{subfigure}
\usepackage[english]{babel}
\usepackage{color}
\begin{document}
\title{\huge Autocorrelation functions: a useful tool for both state and detector characterisation}
\author{Giovanni Chesi}
\affiliation{Department of Science and High Technology, University of Insubria, Via Valleggio 11, I-22100 Como (Italy), E-mail: gchesi@studenti.uninsubria.it}
\author{Alessia Allevi}
\affiliation{Department of Science and High Technology, University of Insubria, Via Valleggio 11, I-22100 Como (Italy), E-mail: alessia.allevi@uninsubria.it}
\author{Maria Bondani}
\affiliation{Institute for Photonics and Nanotechnologies, CNR-IFN, Via Valleggio 11, I-22100 Como (Italy), E-mail: maria.bondani@uninsubria.it}
\date{\today}
\begin{abstract}
{The calculation of autocorrelation functions represents a routinely used tool to characterise quantum states of light. In this paper, we evaluate the $g^{(2)}$ function for detected photons in the case of mesoscopic multi-mode twin-beam states in order to fully investigate their statistical properties starting from measurable quantities. Moreover, we show that the second-order autocorrelation function is also useful to estimate the spurious effects  affecting the employed Silicon-photomultiplier detectors.}
\end{abstract}

\maketitle
\section{Introduction}
The autocorrelation functions, formally introduced by Glauber in 1963 \cite{glauber63}, represents one of the standard tools used to characterise quantum states of light, such as to discriminate between bunched and antibunched light \cite{loudon}. Correlation functions are usually evaluated at the single-photon level: The light state under examination is divided at a balanced beam splitter and the two outputs are detected by means of two avalanche diodes \cite{ave10, elsaesser,somaschi}. We have recently demonstrated that the same scheme can also be adopted to characterise mesoscopic optical states, namely pulsed states containing sizeable numbers of photons in each pulse \cite{antibunching,QMETRO17}. Instead of single-photon detectors, photon-number-resolving (PNR) ones are needed in such a case \cite{JMO,harder16}. For instance, in our work we have employed hybrid photodetectors to prove a behaviour of sub-Poissonian states analogous to antibunching \cite{antibunching}.\\
In the original Glauber's definition, the autocorrelation functions are expressed in terms of normal ordered operators. Nevertheless, in practical situations, analogous definitions expressed in terms of measurable quantities could be desirable \cite{pra12}. Indeed, the link between the autocorrelation functions for photons and those for, e.g., detected photons can be easily found provided the model for the detection process is known.\\ 
Moreover, the calculation of these function can also be useful to extract information about the detectors used to reveal the light under study.\\
In this paper, we emphasize the versatility of the $g^{(2)}$ autocorrelation functions for both the above-mentioned purposes. On the one hand, we show that from the evaluation of autocorrelations we can extract some information about the features of the employed detectors. On the other hand, we prove that correlation functions represent a useful criterion for the characterisation of quantum correlations as well as of nonclassical states.  

\section{Characterisation of Silicon photomultipliers}\label{sec1}
\begin{figure}
\begin{center}
\includegraphics[width=0.7\columnwidth]{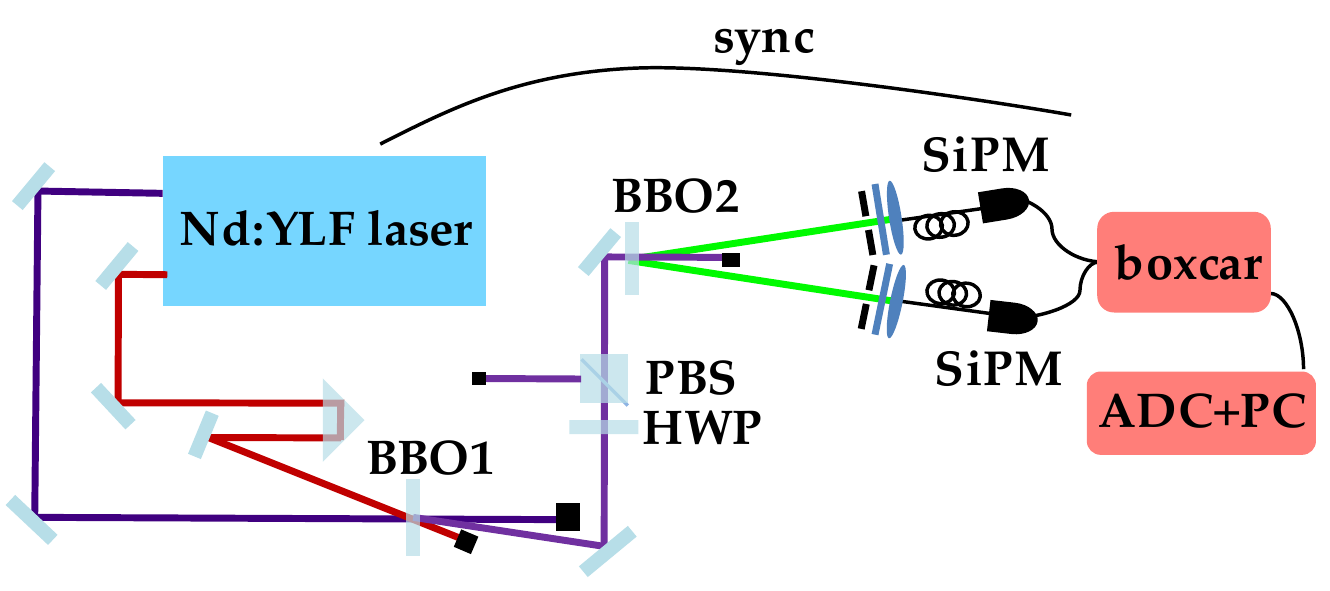}
\end{center}
\caption{Sketch of the experimental setup. See the text for details. }\label{setup}
\end{figure}

\begin{figure}
\begin{center}
\includegraphics[width=0.7\columnwidth]{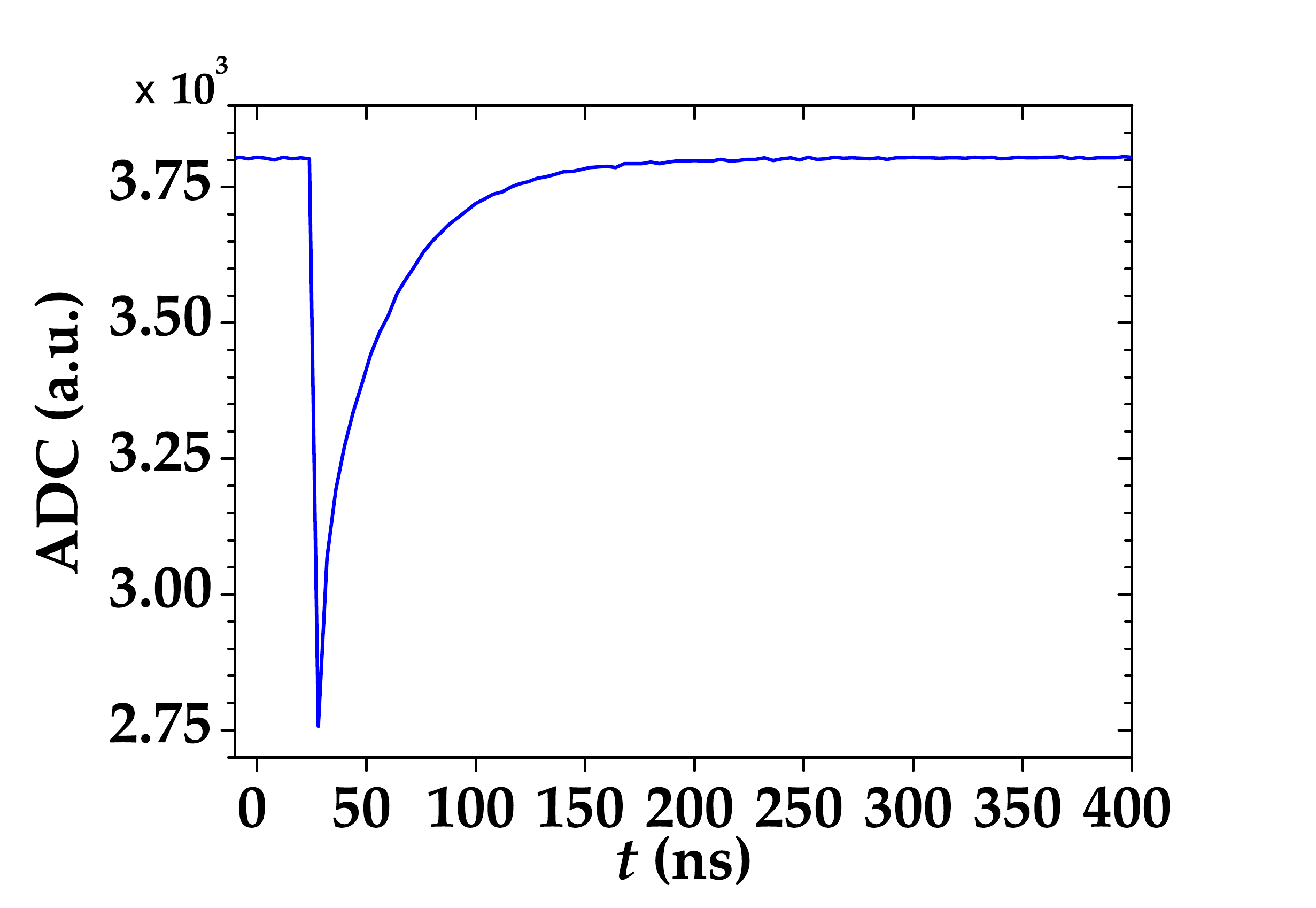}
\end{center}
\caption{Typical single-shot detector signal.}\label{signal}
\end{figure}

As already stated in the Introduction, our measurements have been performed in the mesoscopic intensity regime. In such a case, the $g^{(2)}$ function can be easily evaluated by direct detection of the state under examination. As to the detectors, we decided to employ a commercial class of PNR detectors, namely Silicon photomultipliers (SiPMs).  
These detectors consist of avalanche diodes, called cells, arranged in a matrix of pixels connected to a common output. Every diode is reverse-biased and operates in Geiger-M$\ddot{\rm u}$ller regime \cite{akindinov,bondarenko,saveliev,piemonte,renker}. During the last fifteen years, SiPMs have been employed in many scientific applications, ranging from particle physics experiments to positron emission tomography and biomedical research.
Thanks to their structure, SiPMs are endowed with a good photon-number-resolving capability, which makes them appealing for Quantum Optics, to detect mesoscopic quantum states of light \cite{ramilli}. However, this possibility has been till now prevented by the occurrence of stochastic spurious events, such as dark counts and optical cross-talk \cite{afek,kala12}, as well as by a limited quantum efficiency. 
In the following, we consider the new generation of SiPMs produced by Hamamatsu, in which the cross-talk probability has been lowered and the quantum efficiency increased \cite{hama}.\\ 
According to the model presented in Refs.~\cite{ramilli,chesi}, the response of a SiPM detector can be seen as the convolution of different terms, corresponding to the different detector features.\\ 
First of all, we assume that the detection process is described by a Bernoullian distribution $B_{m, n}(\eta)$. This means that the distribution of detected photons, $P_{\rm el}(m)$, is linked to that of photons, $P_{\rm ph}(n)$, through
\begin{eqnarray}
{\rm P_{el}}(m) &=& \sum_{n=m}^\infty B_{m,n}(\eta) {\rm P_{ph}}(n) \nonumber\\
&=& \sum_{n=m}^\infty {{n}\choose{m}} \eta^m (1- \eta)^{n-m} {\rm P_{ph}}(n),
\end{eqnarray}
where $\eta$ is the detection efficiency, $n$ is the number of incident photons, and $m$ that of detected photons.
Dark counts are spurious avalanches triggered by thermally generated charge carriers. Since they are independent and uncorrelated events, their statistics is Poissonian
\begin{equation}\label{Pdc}
{\rm P_{dc}}(m) = \frac{(\langle m \rangle_{\rm dc})^m}{m!}\exp{(-\langle m \rangle_{\rm dc})},
\end{equation}
where $\langle m \rangle_{\rm dc}$ is the mean value of dark counts.\\ 
Second, we consider optical cross-talk events \cite{gola} that arise when the electrons accelerated during the avalanche process produce brehmsstralung radiation that may trigger avalanches in a neighbouring cell. Hereafter, we assume the cross-talk probability distribution to be \cite{ramilli}
\begin{equation}
C_{k,l}(\epsilon) = {{l}\choose{k-l}} \epsilon^{k-l}(1- \epsilon)^{2l-k},
\end{equation}
where $\epsilon$ is the probability that the avalanche from a cell triggers one neighbour cell, $l$ is the number of photo-triggered avalanches and of dark counts, and $k$ is the resulting number of avalanches including cross talk.\\
Finally, we assume that the amplification (both internal and external) of the detector is described by a multiplicative factor, $\gamma$, so that the single-shot output of the detection chain is $x_{\rm out} = \gamma k$. The distribution of the detector output is given by the convolution of all the previous terms:
\begin{equation}\label{Pout}
{\rm P}(x_{\rm out}) = \gamma \sum_{m=0}^k C_{k,m}(\epsilon) \sum_{j=0}^m {\rm P_{dc}}(j) {\rm P_{el}}(m-j).
\end{equation} 
By using the two moments of the distribution in Eq.~(\ref{Pout}), it is possible to define the $g^{(2)}$ function for the SiPM output as
 \begin{equation}
g^{(2)}( x_{\rm out}) \equiv \frac{\langle x_{\rm out}^2 \rangle}{\langle x_{\rm out} \rangle^2} = \frac{\langle (\gamma k)^2 \rangle}{\langle \gamma k \rangle^2} \equiv g^{(2)}(k) = \frac{\sigma^2(k)}{\langle k \rangle^2} + 1, 
\label{g2out}
\end{equation}
where 
\begin{eqnarray}
\langle k \rangle &=& (1 + \epsilon) (\langle m \rangle + \langle m\rangle_{\rm dc})\\
\sigma^2(k) &=& (1+\epsilon)^2 \left(\sigma^2(m)+\langle m\rangle_{\rm dc}\right) + \epsilon(1+\epsilon)(\langle m \rangle + \langle m\rangle_{\rm dc})\nonumber
\end{eqnarray}
are the mean value and the variance of $k$, respectively. 
Note that $g^{(2)}(k)$ can be also linked to the expression of the autocorrelation function for photons,
\begin{equation}
g^{(2)}(n) = \frac{\langle :n^2: \rangle}{\langle n \rangle^2}.
\end{equation}
In fact, it can be demonstrated that Eq.~(\ref{g2out}) can be re-written as
\begin{equation}
g^{(2)}(k) = 1 + (g^{(2)}(n) -1)\left( 1- \frac{(1+ \epsilon)\langle m \rangle_{\rm dc}}{\langle k \rangle} \right)^2 + \frac{1}{\langle k \rangle} \frac{1+3\epsilon}{1+\epsilon}.
\label{g2new}
\end{equation}
In the Introduction we claimed that the evaluation of the autocorrelation function expressed in terms of measurable quantities can help the determination of the detector features, such as the mean value of dark counts, $\langle m \rangle_{\rm dc}$, and the cross-talk probability, $\epsilon$. Moreover, the expression in Eq.~(\ref{g2new}) also contains information about the light through the term $g^{2}(n)$. For instance, in the case of multi-mode thermal light with $\mu$ modes equally populated $g^{2}(n) = 1 + 1/\mu$.\\
To experimentally prove these statements, we generated a multimode twin-beam (TWB) state by means of parametric down conversion (see Fig.~\ref{setup}). The pump field was the fourth harmonic (at 262 nm, 3.5-ps pulse duration) of a Nd:YLF laser regeneratively amplified at 500 Hz, whose pulses were sent to a $\beta$-barium-borate nonlinear crystal (BBO2, cut angle = 46.7 deg, 6-mm long). 
Two twin portions of TWB states were spatially and spectrally filtered by means of irises and interference filters centered at 523 nm, respectively. The two light components were then delivered to two SiPMs through two multi-mode fibers (600-$\mu$m core diameter). We used two commercial SiPMs (MPPC S13360-1350CS) produced by Hamamatsu \cite{hama1}. Such detectors are endowed with a moderate rate of dark count at room temperature ($\sim$ 140 kHz) and a low cross-talk probability ($\sim 2 \%$). Each detector output was amplified and integrated by means of two synchronous boxcar-gated integrators (SR250, Stanford Research Systems). In particular, we adopted a small value of gate width (10 ns) in order to keep the effects of dark count and delayed cross talk as small as possible. As shown in Fig.~\ref{signal}, the chosen value corresponds to the width of the signal peak.  
The experimental measurements were performed at different values of pump energy, which was modified through a half-wave plate (HWP) followed by a polarizing cube beam splitter (PBS). At each energy value, $10^5$ single-shot acquisitions were performed.\\ 
As declared above, the generation of our TWB state is intrinsically multi-mode \cite{paleari,pra12}. This means that each arm of TWB is described by a multi-mode thermal statistics.
By imposing this condition, Eq.~(\ref{g2new}) reads as 
\begin{equation}\label{g2mthXC}
g^{(2)}(k) = 1 + \frac{1}{\mu} \left(1- \frac{(1+\epsilon) \langle m \rangle_{\rm dc}}{\langle k \rangle} \right)^2 + \frac{1}{\langle k \rangle} \frac{1+3 \epsilon}{1+ \epsilon}.
\end{equation}
Note that, at variance with the autocorrelation function for photons, the maximum value of Eq.~(\ref{g2mthXC}) can be larger than 2.
\begin{figure}
\begin{center}
\includegraphics[width=0.7\columnwidth]{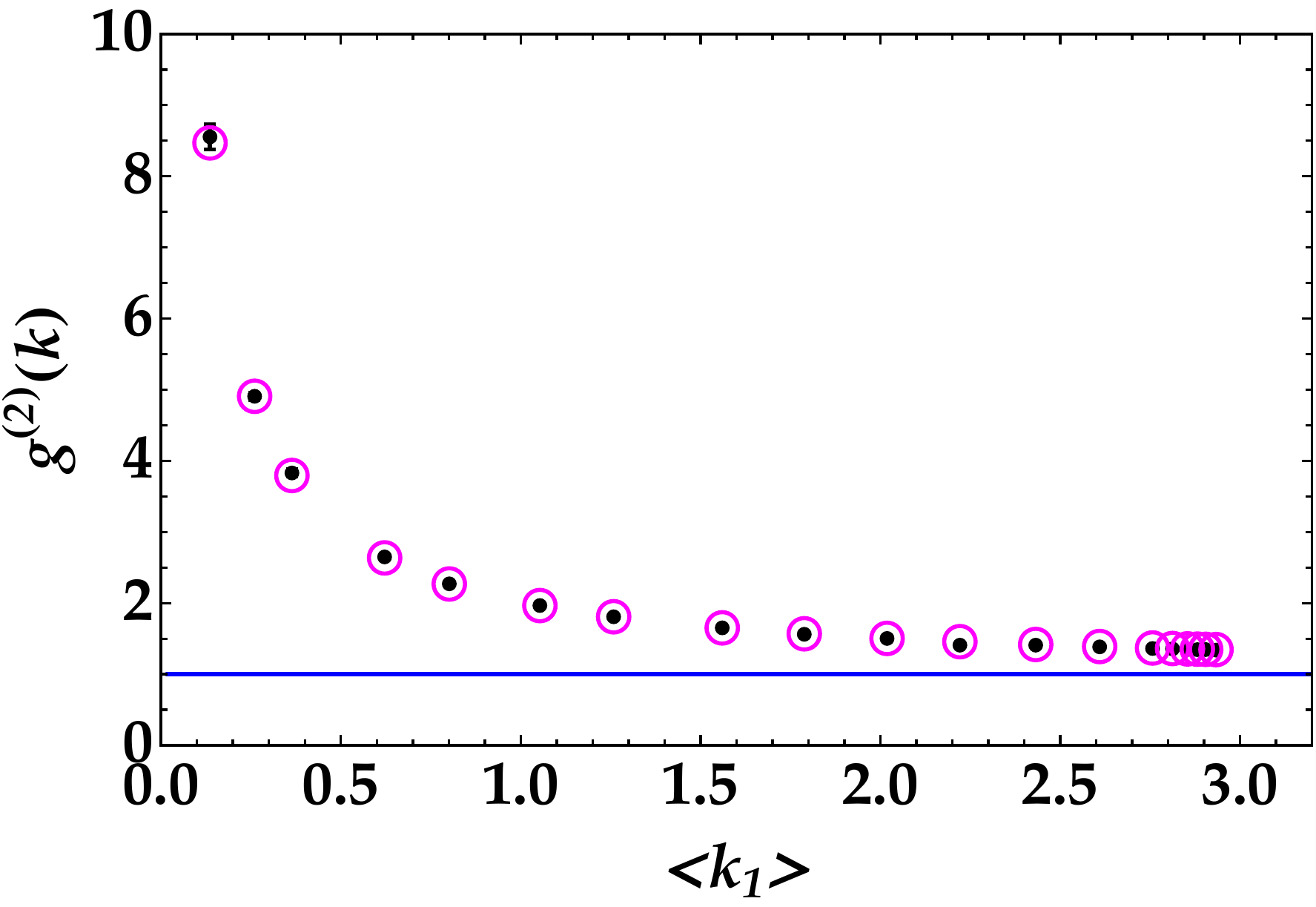}
\includegraphics[width=0.7\columnwidth]{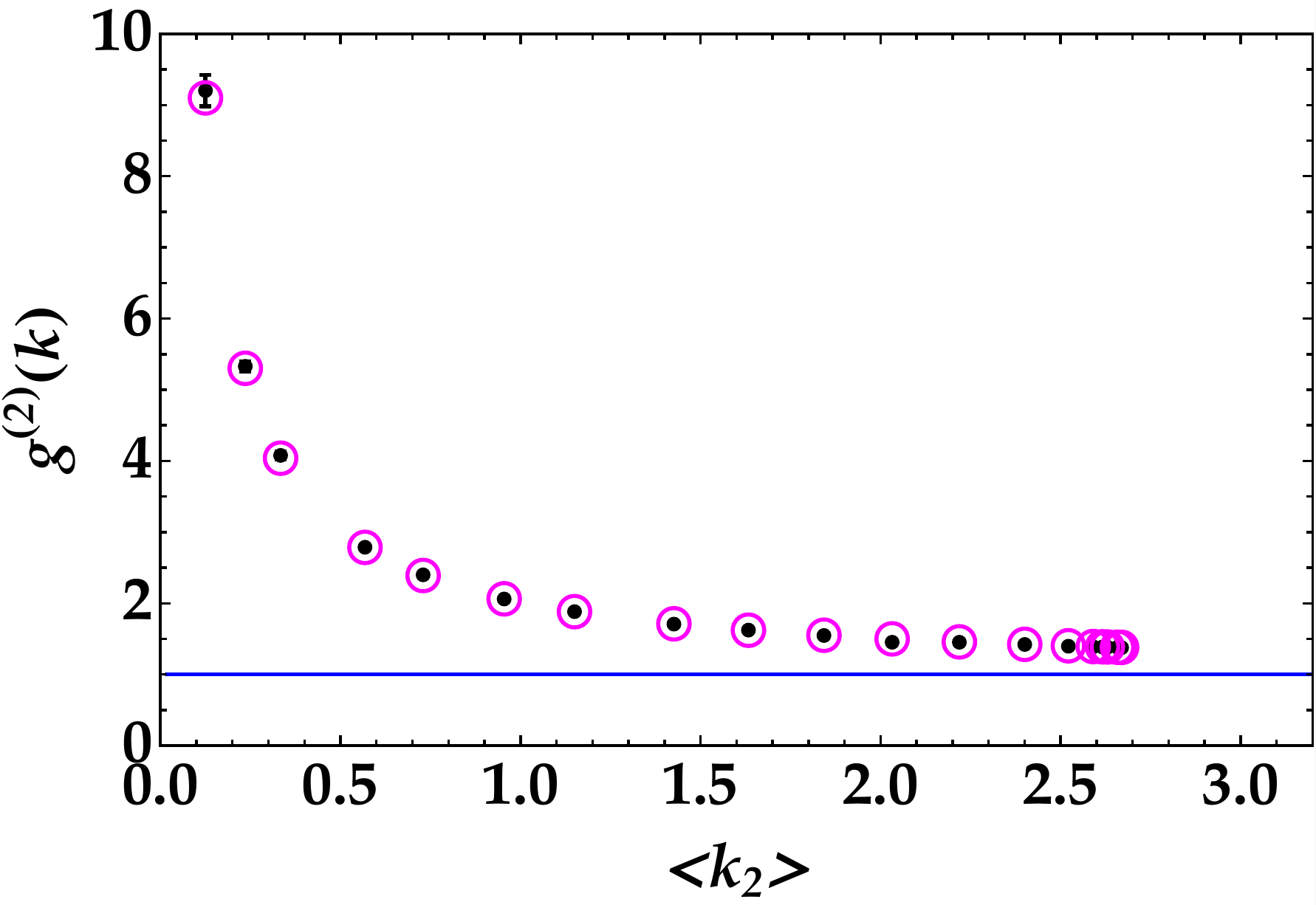}
\end{center}
\caption{Autocorrelation function at different mean numbers of $k$ measured in each TWB arm. The two panels correspond to the two arms. Black dots + error bars: experimenta data; magenta circles: theoretical fitting curve according to Eq.~(\ref{g2mthXC}); blue line: classical boundary. The values of the mean error~\cite{meanerror} with respect to red circles are 0.003 in the upper panel and 0.002 in the lower panel.}\label{g2single}
\end{figure}
In the two panels of Fig.~\ref{g2single}, we plot $g^{(2)}(k)$ as a function of the mean value of $k$ for each TWB arm. The experimental data are shown as black dots + error bars, whereas the theoretical fitting curve (magenta circles) was calculated according to Eq.~(\ref{g2mthXC}), in which we assumed $\mu = 1000$ (reasonable value for a TWB in our experimental conditions \cite{SiPMOL}) and left $\epsilon$ and $\langle m \rangle_{\rm dc}$ as free parameters. In particular, we got: $\epsilon = 0.008$, and $\langle m \rangle_{\rm dc} = 0.001$ in the first arm (upper panel) and $\epsilon = 0.007$, and $\langle m \rangle_{\rm dc} = 0.001$ in the second arm (lower panel). 
The absolute values of $\epsilon$ and $\langle m \rangle_{\rm dc}$ obtained from the fit are quite small. However, in order to quantify their relevance for the calculation of the second-order autocorrelation function, we evaluated the relative variation of $g^{(2)}(k)$ by expanding the function up to the first order of $\alpha$
\begin{equation}\label{deriv}
\frac{\Delta g^{(2)}(k)}{g^{(2)}(k)} = \left. \frac{1}{g^{(2)}(k)|}\right|_{\alpha = 0} \left. \frac{\partial g^{(2)}(k)}{\partial \alpha}\right|_{\alpha = 0} \Delta \alpha \equiv \beta \Delta \alpha,
\end{equation}
where $\alpha$ is either $\epsilon$ or $\langle m \rangle_{\rm dc}$, both evaluated in 0. We note that, for any choice of the parameters, $\beta$ in Eq.~(\ref{deriv}) is always smaller than 1. Thus, since the quantities $\Delta \alpha$ vary in very narrow ranges, the variation of $g^{(2)}(k)$ is negligible.
For this reason, in the following Section we do not consider the contribution of cross talk and dark count in the calculation of the $g^{(2)}$ function connected to other variables.

\section{Characterisation of nonclassical states of light}\label{sec2}
\begin{figure}
\begin{center}
\includegraphics[width=0.7\columnwidth]{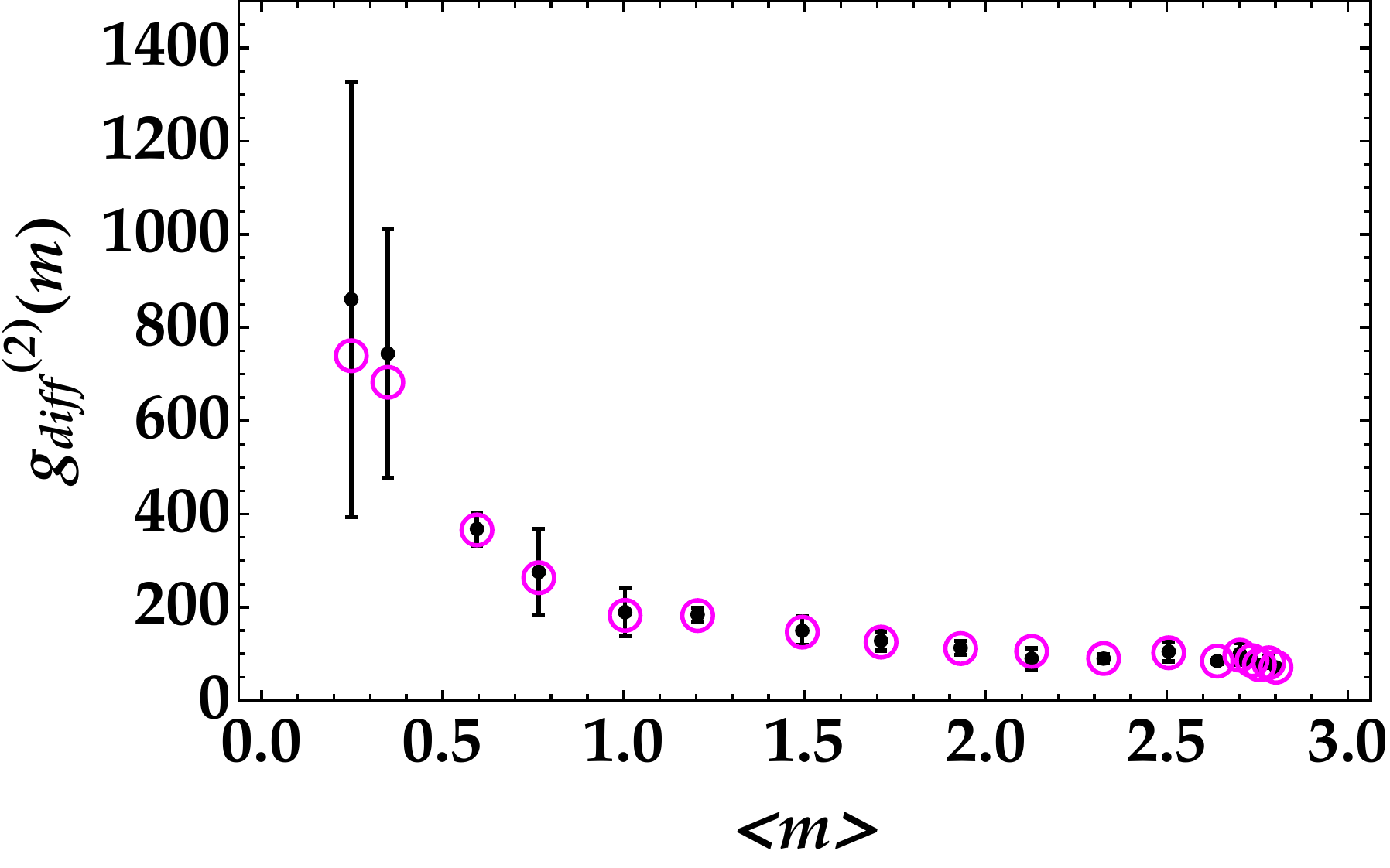}
\end{center}
\caption{Measured $g_{diff}^{(2)}(m)$ as a function of the mean number of photons detected in each TWB arm. Black dots + error bars: experimental data; magenta circles: theoretical expectation according to Eq.~(\ref{g2diffteo}).}\label{g2diff}
\end{figure}

\begin{figure}
\begin{center}
\includegraphics[width=0.7\columnwidth]{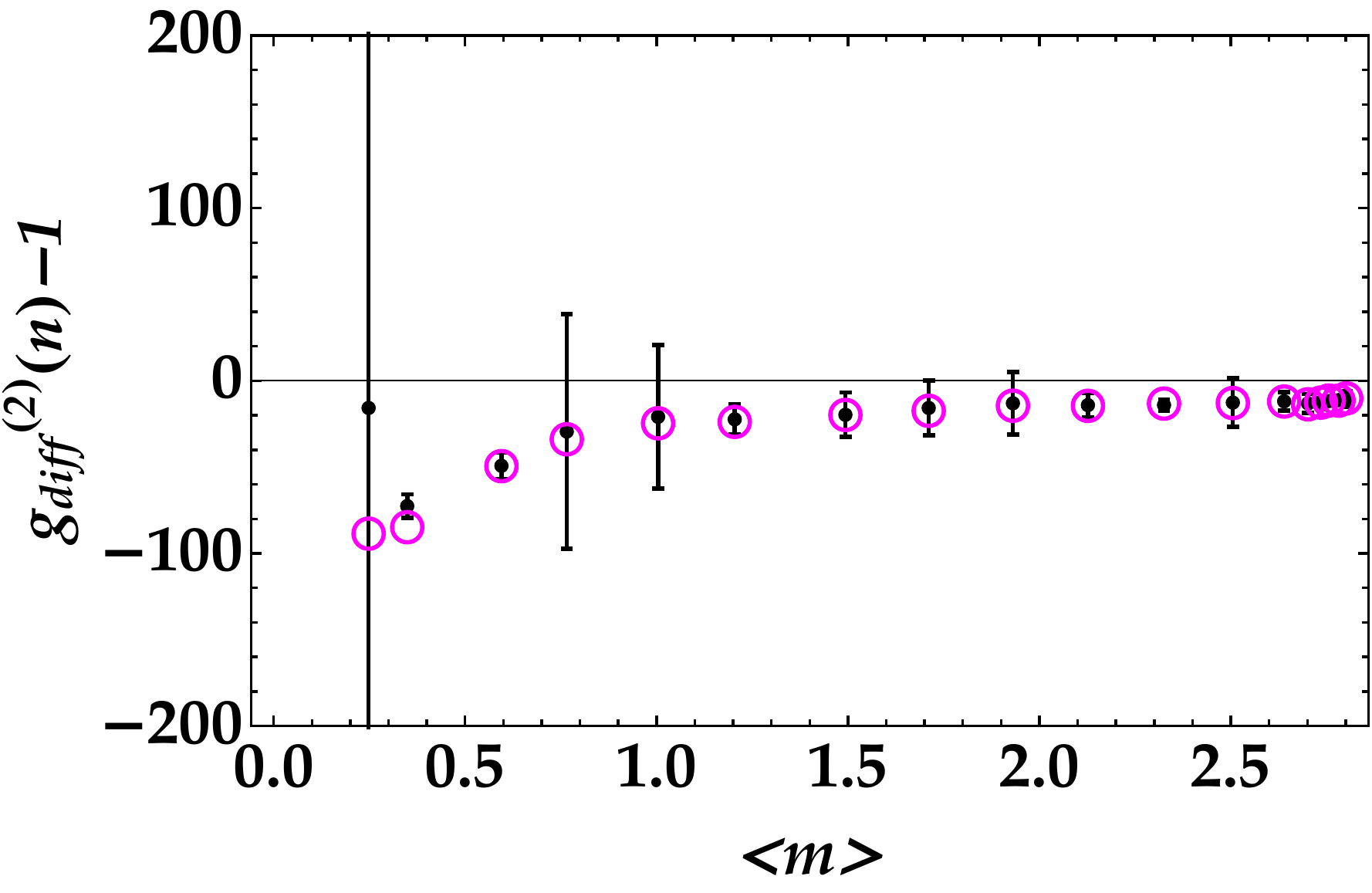}
\end{center}
\caption{Measured $[g_{diff}^{(2)}(n)-1]$ as a function of the mean number of photons detected in each TWB arm. Black dots + error bars: experimental data; magenta circles: theoretical expectation according to Eq.~(\ref{g2diffcrit}).}\label{g2ndiff}
\end{figure}

\begin{figure}
\begin{center}
\includegraphics[width=0.7\columnwidth]{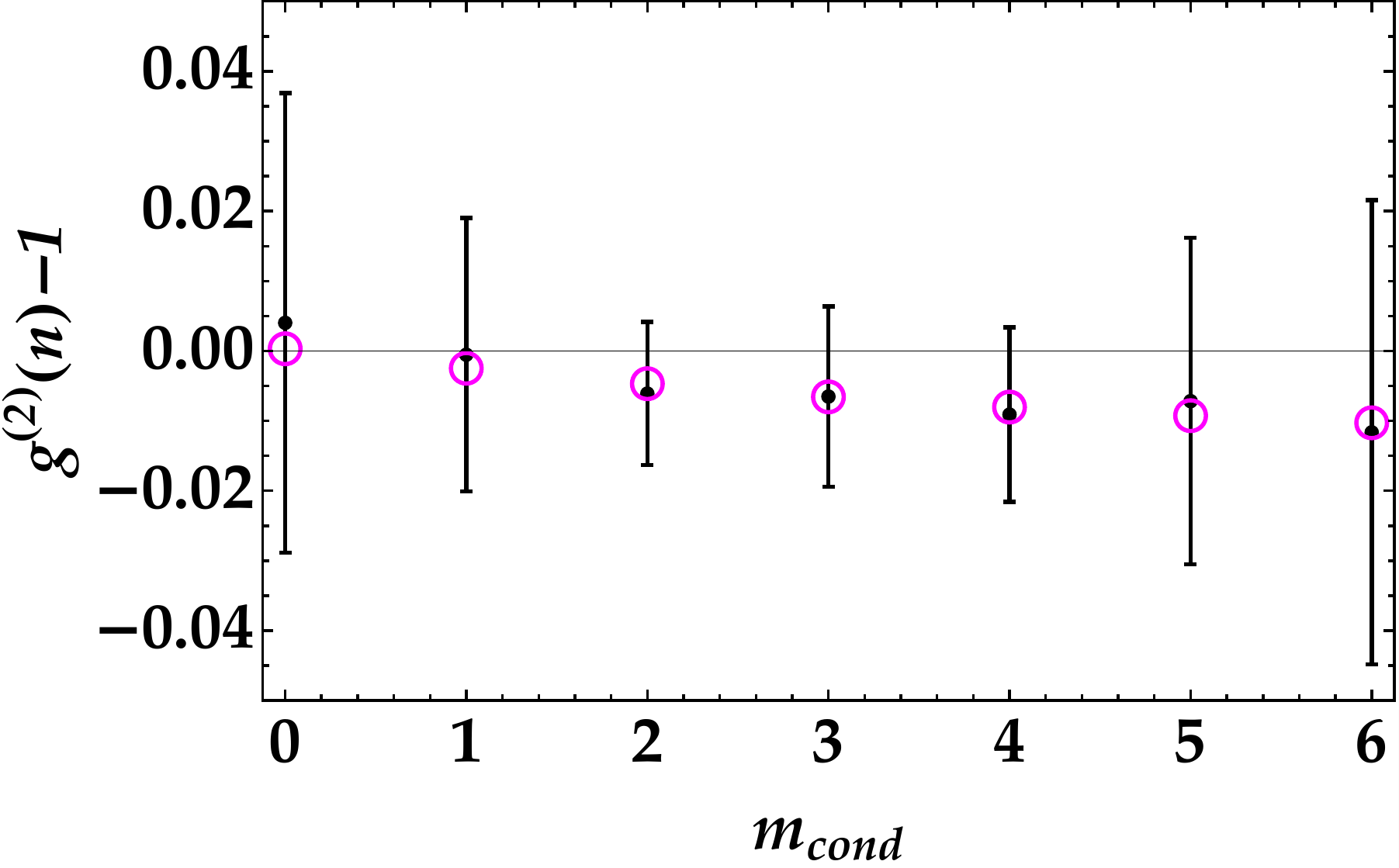}
\end{center}
\caption{Measured values of $[g^{(2)}(n)-1]$ as a function of the conditioning value. Black dots + error bars: experimental data; magenta circles: theoretical expectation according to Eq.~(\ref{eq:g2cond}).}\label{g2cond}
\end{figure}

The TWB states generated as described in Sect.~\ref{sec1} are optical states endowed with nonclassical correlations. This nature can be proved by means of suitable nonclassicality criteria \cite{klyshko96,short83,vogel,arimondo}. Among them, the most used is the noise reduction factor, $R$, which is defined as the ratio between the variance of the photon-number difference between the two twin arms and the shot noise level, namely
\begin{equation}\label{eq:NRF}
R = \frac{\sigma^2(m_1 - m_2)}{\langle m_1 + m_2 \rangle}
\end{equation} 
Values of $R$ lower than 1 testify nonclassicality \cite{pra07}.
To further characterise the TWB states, we evaluated the autocorrelation function for the 
photon-number difference, that is $g_{diff}^{(2)}(m) = \langle (m_1 - m_2)^2 \rangle / \langle m_1 - m_2 \rangle^2$.
We note that this function can be expressed in terms of the noise reduction factor $R$ as
\begin{equation}\label{g2diffteo}
g_{diff}^{(2)}(m) = 1+\frac{R~\langle m_1 + m_2 \rangle}{\langle m_1 - m_2 \rangle^2}.
\end{equation}
In Fig.~\ref{g2diff} we plot the measured values of $g_{diff}^{(2)}(m)$ together with the expectation in Eq.~(\ref{g2diffteo}), in which we used the expression of $R$ for multi-mode thermal TWB states, namely 
\begin{equation} \label{NRFeta}
R = 1 - \frac{2 \sqrt{\eta_1 \eta_2}\sqrt{\langle m_1 \rangle \langle m_2 \rangle}}{\langle m_1 \rangle + \langle m_2 \rangle} + \frac{(\langle m_1 \rangle - \langle m_2 \rangle)^2}{\mu (\langle m_1 \rangle + \langle m_2 \rangle)},
\end{equation}
where $\eta_1$ and $\eta_2$ are the quantum efficiencies in the two arms, calculated in the experimental values of $\langle m_j \rangle$, $\eta_j$ and $\mu$, with $j=1,2$ \cite{lamperti}. 
The experimental data shown in the figure are well superimposed to theory. Note that, since the values of $R$ are within the range (0.87-0.9), the values of $g_{diff}^{(2)}(m)$  are in general quite large, no matter the absolute values of $\langle m_1 \rangle$ and $\langle m_2 \rangle$. This fact emphasizes that the term $\langle m_1 - m_2 \rangle^2$ is quite small, thus proving a good balancing between the light detected in the two arms.\\
While the evaluation of $g_{diff}^{(2)}(m)$ is useful to investigate the balancing between the numbers of photons detected in the two arms, the calculation of the analogous expression for photons, $i.e.$ $g_{diff}^{(2)}(n)$ can give information about the quantum nature of photon-number correlations.
By definition
\begin{equation}\label{g2nodiff}
g_{diff}^{(2)}(n) = \frac{\langle:(n_1 - n_2)^2:\rangle}{\langle n_1 -n_2 \rangle^2} = \frac{\langle(n_1 - n_2)^2\rangle}{\langle n_1 -n_2 \rangle^2} - \frac{\langle n_1 + n_2\rangle}{\langle n_1 -n_2 \rangle^2}.
\end{equation}
Assuming that the detection efficiency in the two arms of TWB is the same, $\eta_1 = \eta_2 = \eta$, it is possible to demonstrate that the relation between $g_{diff}^{(2)}(n)$ and $g_{diff}^{(2)}(m)$ reads as follows:
\begin{equation}\label{g2difflink}
g_{diff}^{(2)}(n) = g_{diff}^{(2)}(m) - \frac{\langle m_1 + m_2\rangle}{\langle m_1 -m_2 \rangle^2}.
\end{equation}
By using Eq.~(\ref{g2diffteo}) in Eq.~(\ref{g2difflink}), the autocorrelation function $g_{diff}^{(2)}(n)$ can be directly connected to the noise reduction factor
\begin{equation}\label{g2diffcrit}
g_{diff}^{(2)}(n) -1 = (R-1)\frac{\langle m_1 + m_2\rangle}{\langle m_1 -m_2 \rangle^2}.
\end{equation}
According to Eq.~(\ref{g2diffcrit}), the negativity of the quantity $[g_{diff}^{(2)}(n)-1]$ can be used as nonclassicality criterion as an alternative to $R<1$.  
In Fig.~\ref{g2ndiff} we show the values of $[g_{diff}^{(2)}(n) -1]$ calculated from the experimental data as black dots + error bars. In the same figure we also plot the theoretical expectation according to Eq.~(\ref{g2diffcrit}), in which $R$ was evaluated according to Eq.~(\ref{NRFeta}). The negativity of the plotted quantity proves the quantum nature of photon-number correlations.\\ 
Since all the measured TWB states are nonclassically correlated, the employed SiPMs can be used to perform multi-photon conditioning operations in order to produce sub-Poissonian states of light \cite{laurat03,laurat04,ourj06,Iskhakov16}. Indeed, we have already demonstrated that, when a certain number of photons is selected in one TWB arm, the distribution of photons in the other arm is narrower than a Poissonian distribution \cite{lamperti}. This fact can be quantified either by calculating the Fano factor, $F(m) = \sigma^2_m/\langle m \rangle$, of the obtained photon-number statistics or evaluating the $g^{(2)}$ function for photons.
In fact, it is possible to demonstrate \cite{antibunching} that 
\begin{equation}	 \label{eq:g2cond}
g^{(2)}(n) - 1 = \frac{F(m)-1}{\langle m \rangle}.
\end{equation} 
In the case of conditional states obtained by multi-mode thermal TWB states, the expression of $F(m)$ is not trivial, but it is analytic
\begin{eqnarray}
\label{eq:Fanocond}
\hspace{-0.1cm}F(m) \hspace{-0.3cm}&=& \hspace{-0.3cm}(1- \eta)+ (1- \eta) \cdot\\
&\cdot & \hspace{-0.3cm} \frac{\langle m \rangle(m_{\rm cond}+\mu)(\langle m \rangle+\eta \mu)}{(\langle m \rangle+ \mu)
[(m_{\rm cond}+\mu)(\langle m \rangle+\eta \mu)-\eta \mu (\langle m \rangle+\mu)]} ,\nonumber
\end{eqnarray}
where $\eta$ is the overall detection efficiency, $\langle m \rangle$ the mean value of the unconditioned state, and $m_{\rm cond}$ the conditioning value, that is the value of photons measured in one arm according to which the values of the other arm are selected.\\
In Fig.~\ref{g2cond} we plot the experimental values of $[g^{(2)}(n) -1]$ (black dots + error bars) as a function of different conditioning values. Even if the negativity of the plotted quantity is not so large, it is sufficient to prove that the produced states are nonclassical. In the same figure, we also show the theoretical expectations (magenta circles) according to Eq.~(\ref{eq:g2cond}), in which Eq.(\ref{eq:Fanocond}) was used. As in the case of the $g^{(2)}$ function for the photon-number difference, also for the Fano factor we used the experimental values of $\eta$, $\mu$ and $\langle m \rangle$. In particular, for the data in the figure $\langle m \rangle = 2.64$. We notice that the data are well superimposed to theory.   

\section{Conclusions}
In conclusions, we have shown the twofold usefulness of the $g^{(2)}$ autocorrelation function written in terms of measurable quantities. On the one hand, we have proved that its evaluation can be used to characterise the main features of the employed detector. In particular, the spurious stochastic features that affect SiPMs can be easily determined from the expression of the second-order autocorrelation function. Moreover, for the specific choice of operational detector parameters and of the integration gate width used to acquire the detector output we have proved that such effects are substantially negligible. On the other hand, we have used the $g^{(2)}$ function to fully investigate the statistical properties and the correlated nature of mesoscopic multi-mode thermal TWB states. In more detail, we have evaluated the autocorrelation function of the photon-number difference and demonstrated that it can be used as a nonclassicality criterion for correlations. Furthermore, we have exploited the quantum nature of such TWB states to produce sub-Poissonian conditional states in post-selection, which exhibit the negativity of the quantity $[g^{(2)}(n)-1]$ for different conditioning values.
The good quality of the experimental results and their good agreement with the theoretical expectations encourage the further exploitation of SiPMs in the context of Quantum Optics, especially to improve the investigation of real mesoscopic states of light.


\begin{thebibliography}{99}
\bibitem{meanerror} The mean error is defined as $\sum_{i=1}^N |g^{(2)}_{\rm exp} - g^{(2)}_{\rm th}|/(N-p)$, where exp and th are referred to data and theory, respectively and $p$ is the number of used fitting parameters.
\end{thebibliography}
\end{document}